\documentclass{iaus}
\usepackage{graphics}

\newcommand{\teff}{T$_{\rm eff}$ }

\title[Li depletion in the Spite plateau] 
{Observational signatures for depletion in the Spite plateau:
solving the cosmological Li discrepancy?}

\author[J. Mel\'endez, L. Casagrande, I. Ram\'{i}rez, M. Asplund  \& W. J. Schuster]   
{Jorge Mel\'endez$^1$,
Luca Casagrande$^2$,
Iv\'an Ram\'{i}rez$^2$,
Martin Asplund$^2$
\and William J. Schuster$^3$
 }

\affiliation{$^1$Centro de Astrof\'{i}sica, Universidade do Porto, Rua das Estrelas, 4150-762 Porto, Portugal \\
email: {\tt jorge@astro.up.pt} 
\\[\affilskip]
$^2$Max-Planck-Institut f\"ur Astrophysik, 
Karl-Schwarzschild-Str. 1, Postfach 1317, D-85741 Garching, 
Germany
\\[\affilskip]
$^3$Observatorio Astron\'omico Nacional, UNAM, 
Apartado Postal 877, Ensenada, BC, CP 22800, Mexico
}

\pubyear{2010}
\volume{268}  
\jname{Light elements in the Universe}
\editors{C. Charbonnel, M. Tosi, F. Primas \& C. Chiappini, eds.}
\begin{document}

\maketitle

\begin{abstract}
We present Li abundances for 73 stars in the metallicity
range -3.5 $<$ [Fe/H]  $<$ -1.0 using improved IRFM temperatures (Casagrande et al. 2010)
with precise 
\emph{E(B-V)} values obtained mostly from interstellar NaI D lines, and
high-quality equivalent widths ($\sigma_{EW}$
$\sim 3\%$). 
At all metallicities we uncover a fine-structure in the Li abundances of Spite
plateau stars, which we trace to Li depletion that depends on both metallicity and mass.
Models including atomic diffusion and turbulent mixing seem to reproduce the observed Li depletion assuming a primordial Li abundance $A_{\rm Li}$  = 2.64 dex 
(MARCS models) or
2.72 (Kurucz overshooting models),
in good agreement with current predictions ($A_{\rm Li}$ = 2.72) from standard
BBN. We are currently expanding our sample to have a better coverage of
different evolutionary stages at the high and low metallicity ends,
in order to verify our findings.

\end{abstract}

\firstsection 

\section{Introduction}

One of the most important discoveries in the study of
the chemical composition of stars was made in 1982
by M. and F. Spite, who found an essentially constant
Li abundance in warm metal-poor stars (Spite \& Spite 1982), 
a result interpreted as a relic of
primordial nucleosynthesis. Due to its cosmological
significance, there have been many studies
devoted to Li in metal-poor field stars
(e.g. Mel\'endez \& Ram\'{i}rez 2004; Boesgaard et al. 2005;
Charbonnel \& Primas 2005; Nissen et al. 2005; 
Asplund et al. 2006; Bonifacio et al. 2007; Shi et al. 2007;
Hosford et al. 2009; Aoki et al. 2009),
with observed Li abundances at the lowest [Fe/H] 
from as low as $A_{\rm Li}$  = 1.94 to as high as $A_{\rm Li}$  = 2.37.

Using the theory  of big bang nucleosynthesis (BBN) and
the baryon density obtained from WMAP data,
a primordial Li abundance
of $A_{\rm Li}$  = 2.72$_{-0.06}^{+0.05}$ is predicted 
(Cyburt et al. 2008),
which is
a factor of 2-6 times higher than the Li abundance inferred
from halo stars.
There have been many theoretical studies on non-standard
BBN trying to explain the cosmological Li discrepancy by
exploring the frontiers of new physics
(e.g. Coc et al. 2009; Jedamzik 
\& Pospelov 2009; Kohri \& Santoso 2009).
Alternatively, the Li problem 
could be explained by a reduction of the original Li stellar abundance
due to internal processes (i.e., by stellar depletion).
In particular, stellar models including atomic diffusion and mixing can 
deplete a significant fraction of the initial 
Li content (e.g., Richard et al. 2005; Korn et al. 2006; Lind et al. 2009).

Due to the uncertainties in the Li abundances
and to the limited samples available, only limited comparisons of 
models of Li depletion with stars in a broad range of mass and
metallicities  have been performed. 
We are performing such a study (Mel\'endez et al. 2010), achieving errors 
in Li abundance lower than 0.035 dex, for a large sample 
of metal-poor stars (-3.5 $<$ [Fe/H] $<$ -1.0), for
the first time with precisely determined \teff 
(Casagrande et al. 2010) and masses in a 
relatively broad mass range (0.6-0.9 M$_\odot$).

Our new temperature scale (Casagrande et al. 2010) is highly accurate, since it has been
calibrated using solar twins (Mel\'endez et al. 2009; Ram\'{i}rez et al. 2009),
and it has also been tested using stellar diameters and absolute flux spectra
(Casagrande et al. 2010).
Our temperatures are also very precise because for most stars the interstellar (IS) NaI D lines
were used in the determination of E(B-V) (see e.g. Ram\'{i}rez et al. 2006),
implying thus in relatively low errors in our IRFM effective temperatures.
An example of a nearby halo star (HD 140283) showing no detectable
IS NaD lines (E(B-V) = 0.00) is presented in Fig. 1.

\begin{figure}
\centering
\resizebox{10cm}{!}{\includegraphics*{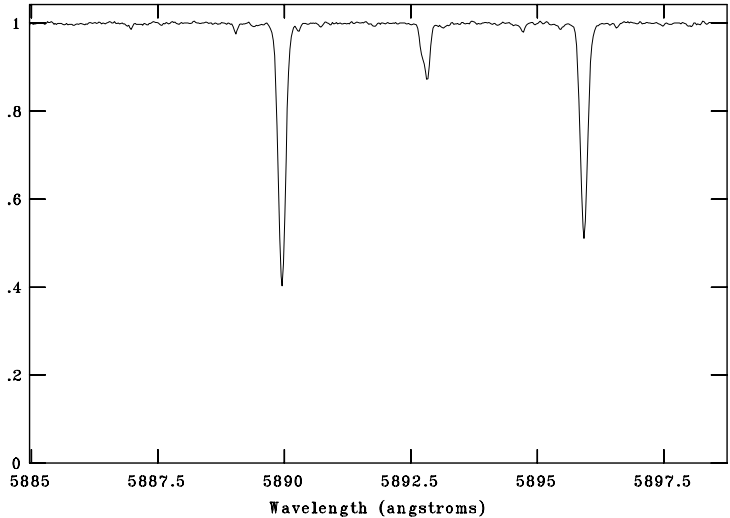} }
\caption{Keck HIRES spectrum around the NaD lines 
for the nearby (57 pc) metal-poor ([Fe/H] = $-$2.4)
star HD 140283. The only stellar lines that can be easily seen are NaD (the two strongest features) 
and a blend of NiI/FeI around 589.3nm. The other features are mostly due to 
weak telluric water vapor lines. 
From the absence of IS NaD lines we infer E(B-V) = 0.000 for this star, although
small amounts of reddening (at the level of 0.001 magnitudes) can not be excluded
}
\label{figteff}
\end{figure}

\begin{figure}
\centering
\resizebox{12cm}{!}{\includegraphics{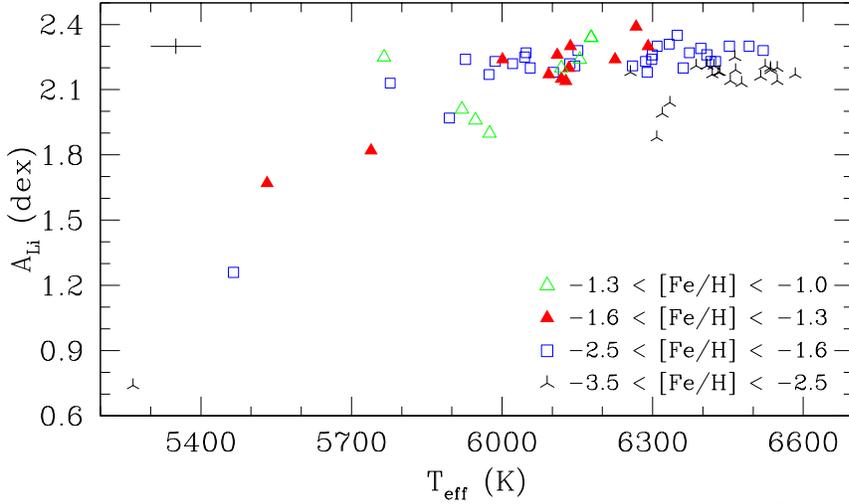} }
 \caption{Li abundances vs. \teff for our sample of metal-poor stars in different metallicity
ranges. The spread at any given metallicity is much larger than the
error bar. Figure taken from \cite{mel10}
}
\label{figteff}
\end{figure}

\begin{figure}
\centering
\resizebox{10cm}{!}{\includegraphics{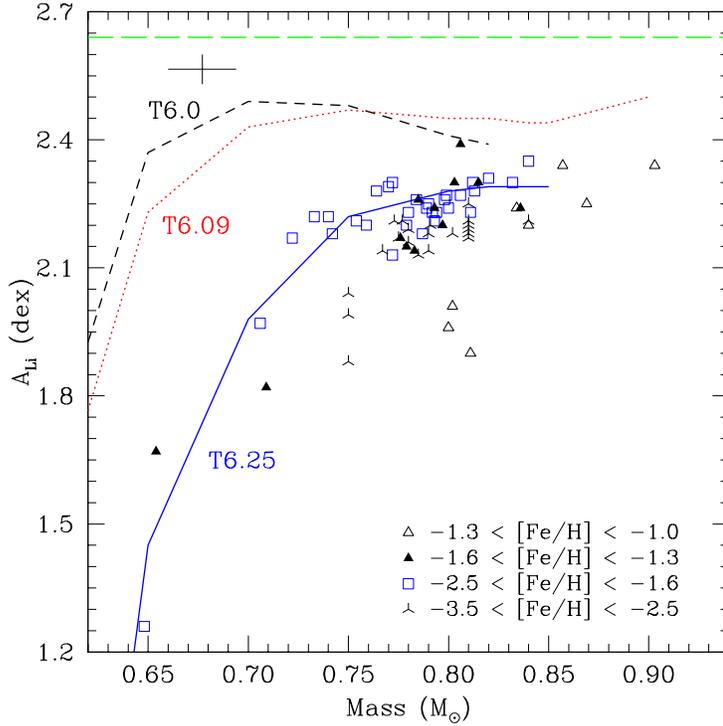} }
 \caption{Li abundances as a function of stellar mass in different
metallicity ranges. Models at [Fe/H] = $-2.3$ including diffusion and
T6.0 (short dashed line), T6.09 (dotted line) 
and T6.25 (solid line) turbulence (Richard et al. 2005) are shown. 
The models have been rescaled to an initial $A_{\rm Li}$=2.64 (long dashed line).
Figure taken from \cite{mel10}
}
   \label{figura}
\end{figure}

\section{Li depletion in Spite plateau stars}

Our work shows that Li is depleted in Spite plateau stars (Fig. 2).
The spread of the Spite plateau at any metallicity is much larger 
than the error bar, as can be clearly seen in Fig. 2. 
Also, there is a correlation between Li and stellar mass 
at any probed metallicity (Fig. 3), showing thus that Li has been depleted
in Spite plateau stars at any metallicity. 
In Fig. 3 we confront the 
stellar evolution predictions of \cite{ric05} with our inferred
stellar masses and Li abundances. 
The models include the effects of atomic diffusion, 
radiative acceleration and gravitational settling
but moderated by a parametrized turbulent mixing.
The agreement is very good when adopting a turbulent model of T6.25 
and an initial $A_{\rm Li}  = 2.64$. The stellar NLTE Li abundances used 
above were obtained with the latest MARCS models (Gustafsson et al. 2008), 
but if we use instead the Kurucz convective overshooting
models, then the required initial abundance to explain our data 
would be $A_{\rm Li} = 2.72$.

Our results imply that the Li abundances observed in Li plateau
stars have been depleted from their original values and therefore
do not represent the primordial Li abundance
(see also Korn et al. 2006 and Lind et al. 2009 for additional
signatures of Li depletion in stars of the globular cluster NGC 6397).
It appears that the observed Li abundances in metal-poor stars
can be reasonably well reconciled with the predictions 
from standard Big Bang nucleosynthesis (e.g. Cyburt et al. 2008)
by means of more realistic stellar evolution models that
include Li depletion through diffusion and turbulent mixing
(Richard et al. 2005).
We caution however, that, although encouraging, our results 
should not be viewed as proof of the correctness of the 
Richard et al. models until the
free parameters required for the stellar modeling are 
better understood from basic physical principles.
In this context, new physics should not be discarded yet
as a solution of the cosmological Li discrepancy, 
as perhaps the low Li-7 abundances in metal-poor stars 
might be a signature of supersymmetric particles in the early universe,
which could also explain the Li-6 detections in
metal-poor stars (e.g. Asplund et al. 2006; Asplund \& Mel\'endez 2008).

We are expanding our sample to have a better coverage of different 
evolutionary stages at all metallicities. Our expanded sample 
(Mel\'endez et al. 2010) will allow us to verify if the Li plateau is indeed depleted
at low and high metallicities.



\end{document}